\documentclass{moriond}

\def\Journal#1#2#3#4{{#1} {\bf #2}, #3 (#4)}

\def\PRL{\em Phys. Rev. Lett.}
\def\PRD{{\em Phys. Rev.} D}

\def\be{\begin{equation}}
\def\ee{\end{equation}}
\def\bea{\begin{eqnarray}}
\def\eea{\end{eqnarray}}

\newcommand{\Photo}{}

\begin{document}
\vspace*{4cm}
\title{Dark matter Axion search with riNg Cavity Experiment\\
DANCE: Development of control system for long-term measurement}

\author{Hiroki Fujimoto$^1$, Yuka Oshima$^1$, Masaki Ando$^1$,Tomohiro Fujita$^2$,\\
Yuta Michimura$^1$, Koji Nagano$^3$, and Ippei Obata$^4$}

\address{$^1$Department of Physics, University of Tokyo, Bunkyo, Tokyo 113-0033, Japan\\
$^2$Institute for Cosmic Ray Research, University of Tokyo, Kashiwa 277-8582, Japan\\
$^3$Institute of Space and Astronautical Science, Japan Aerospace Exploration Agency, Sagamihara, Kanagawa 252-5210, Japan\\
$^4$Max-Planck-Institut für Astrophysik, Karl-Schwarzschild-Straße 1, 85741 Garching, Germany}

\maketitle\abstracts{
Axion-like particles (ALPs) are pseudo-scalar particles that are candidates for ultralight dark matter. ALPs interact with photons slightly and cause the rotational oscillation of linear polarization.
DANCE searches for ALP dark matter by enhancing the rotational oscillation in a bow-tie ring cavity.
	The signal to noise ratio of DANCE can be improved by long-term observation, and we are planning a year-long observation for the final DANCE. In this document, I will report on the control systems of the ring cavity we developed for the future long-term observation.}

\section{Introduction}

Researches on cosmology and astrophysics have revealed that more than 80\% of the matter in the universe consists of dark matter. The nature of dark matter is still unknown and there has been many searches for various dark matter candidates.
”Dark matter Axion search with riNg Cavity Experiment” (DANCE) \cite{DANCE} is a laser interferometric detector of axion-like particles (ALPs), which are the candidates of ultralight dark matter.
ALPs are pseudo-scalar particles predicted from high energy physics such as string theory and behave as classical coherent fields due to their small mass, $m_a\ll$ eV.

ALP produced via the misalignment mechanism behaves as dark matter by oscillating its background field \cite{JP}$^,$~\cite{LF}$^,$~\cite{MD}$^,$~\cite{PA}.
The presence of the oscillating ALP field causes a small phase velocity difference between left- and right-handed circularly polarized light \cite{SM1}$^,$~\cite{SM2}. From the point of view of linear polarization, this phase velocity difference can be regarded as the rotational oscillation of the polarization plane, and the signal of ALP appears as the sidebands with polarization orthogonal to the carrier polarization \cite{HL}.
In the DANCE experiment, this rotational amplitude induced by the ALP field can be amplified by using an optical ring cavity composed by 4 mirrors (see Fig. 1) and extending the effective light path length. 

The rotational oscillation of linear polarization can be regarded as coherent during $\tau$, during which the cavity of DANCE is in the de Broglie wavelength of the ALP dark matter. With the observation for $T_{\rm{obs}}$, the signal to noise ratio (SNR) of DANCE can be improved as \cite{DB}
\[
  \rm{SNR} \propto \left\{ \begin{array}{ll}
    T_{\rm{obs}}^{1/2} & (T_{\rm{obs}} < \tau) \\
    T_{\rm{obs}}^{1/4}  & (T_{\rm{obs}} > \tau), 
  \end{array} \right.
   \quad \tau \simeq 1 \, \rm{yr} \left( \frac{10^{-16} \, \rm{eV}} {\it{m_a}} \right).
\]
Therefore, the long-term observation (e.g. 1 year) is important for DANCE to search ALPs widely.

However, there had been some issues in the prototype experiment of DANCE (DANCE Act-1) for long-term observation.
Firstly, in the previous setup \cite{DANCE2}, the cavity was locked to resonance by only one piezo actuator, and the lock continued for only one hour due to the narrow actuation range of the actuator. 
Secondly, when the cavity drops out of lock, we needed to lock the cavity manually again.

In this document, we report the double-loop feedback control system and the automated cavity locking system developed to solve these issues for long-term observation of DANCE Act-1. 
By applying the double-loop feedback control, the lock duration time has been improved to more than 60 hours, and by using the automated cavity locking system, the cavity resonance has become capable of recovering from the unlock automatically.

\section{Method}

\begin{figure}
\centerline{\includegraphics[width=0.6\linewidth]{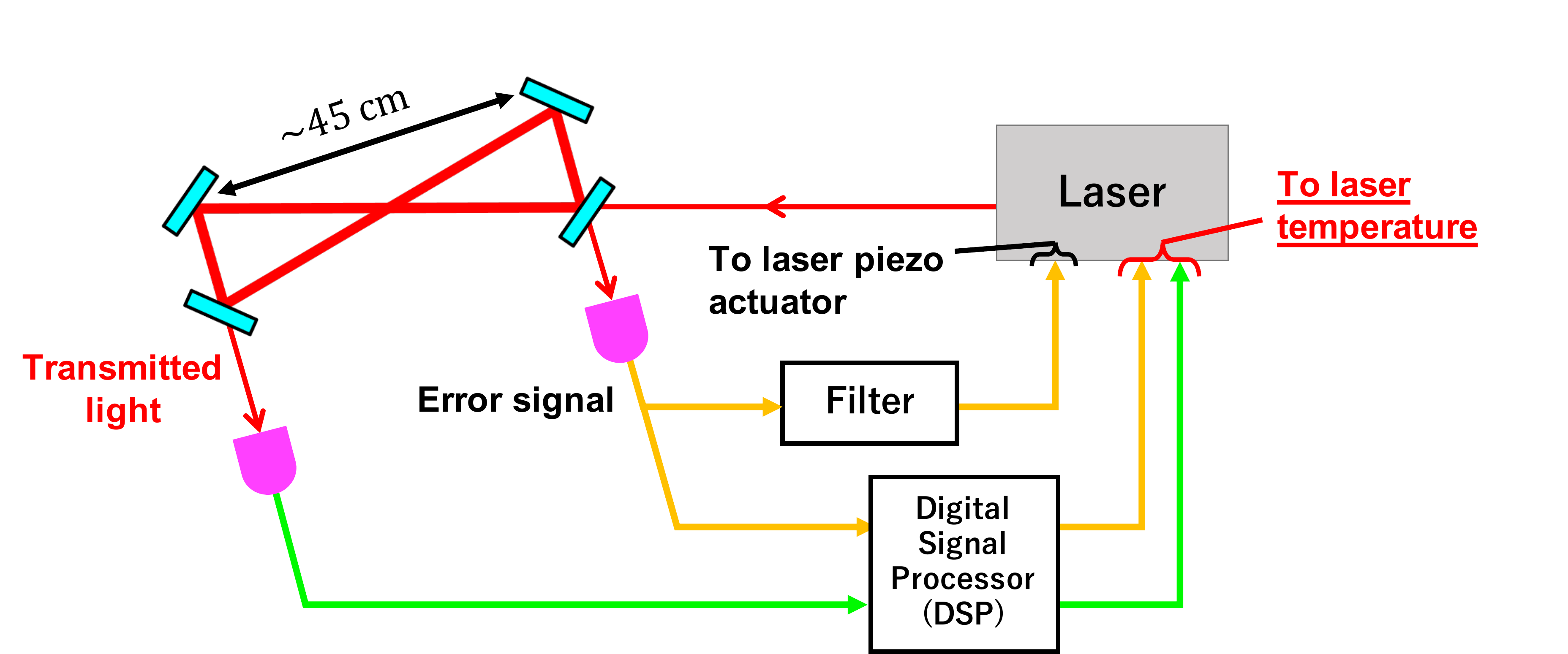}}
\caption[]{Configuration of the double-loop feedback control system (yellow) and the automated cavity locking system (green).}
\label{fig:radish}
\end{figure}

\subsection{Double-loop feedback control system}

Generally, a laser frequency can be locked to an optical cavity resonance by the feedback control of the laser frequency or the cavity length.
In the previous setup of DANCE Act-1, we locked the cavity by using the piezo actuator in the laser source.
However, the piezo actuator has a narrow actuation range of $51\pm12\, \rm{MHz}$, and  the cavity was easily unlocked within about one hour due to low frequency disturbances such as temperature changes.

To improve the lock duration time, we developed the double-loop feedback control system by using the temperature of the laser crystal as the second actuator (Fig. 1). The laser frequency can be changed also by the thermal expansion of the laser crystal.
The actuation range of this temperature of the laser source is $5.46\pm0.12\, \rm{GHz}$ and much wider than that of piezo actuator.
Because the temperature in the laser source can takes on the low frequency disturbance instead of the piezo actuator by the double-loop feedback configuration, the drift in the feedback signal to the piezo actuator can be suppressed, preventing the feedback signal from going out of the actuation range.
To implement this system, we used SEAGULL mini by MTT Corporation as a digital signal processor (DSP) to design a filter for the second actuator.

\subsection{Automated cavity locking system}

In order to lock the cavity to the resonance, the laser frequency needs to be roughly adjusted to the resonant frequency at first. After that, the feedback control system becomes capable of locking the cavity because the error signal can be obtained around the resonant frequency.
This rough adjustment is also needed when the cavity is unlocked due to the temperature change or sudden shock and vibration.
In the previous setup of DANCE Act-1, we needed to do this adjustment manually, although the unlocking of the cavity in the long-term observation is inevitable.

To enable long-term observation, we developed the automated cavity locking system, which takes on the rough adjustment of the laser frequency automatically.
The green line in Fig. 1 shows  the scheme of the automated cavity locking system.
To implement this system, the DSP is used again. The DSP monitors the transmitting light from the cavity and identifies whether the cavity is locked or unlocked. If the cavity is unlocked, the DSP sweeps the temperature of the laser source until the cavity is locked again.
The sweep direction is always set upward in this experiment.

\section{Results and Discussion}

\subsection{Lock duration time measurement}

\begin{figure}

\begin{minipage}{0.33\linewidth}
\centering
\includegraphics[width=1.35\linewidth]{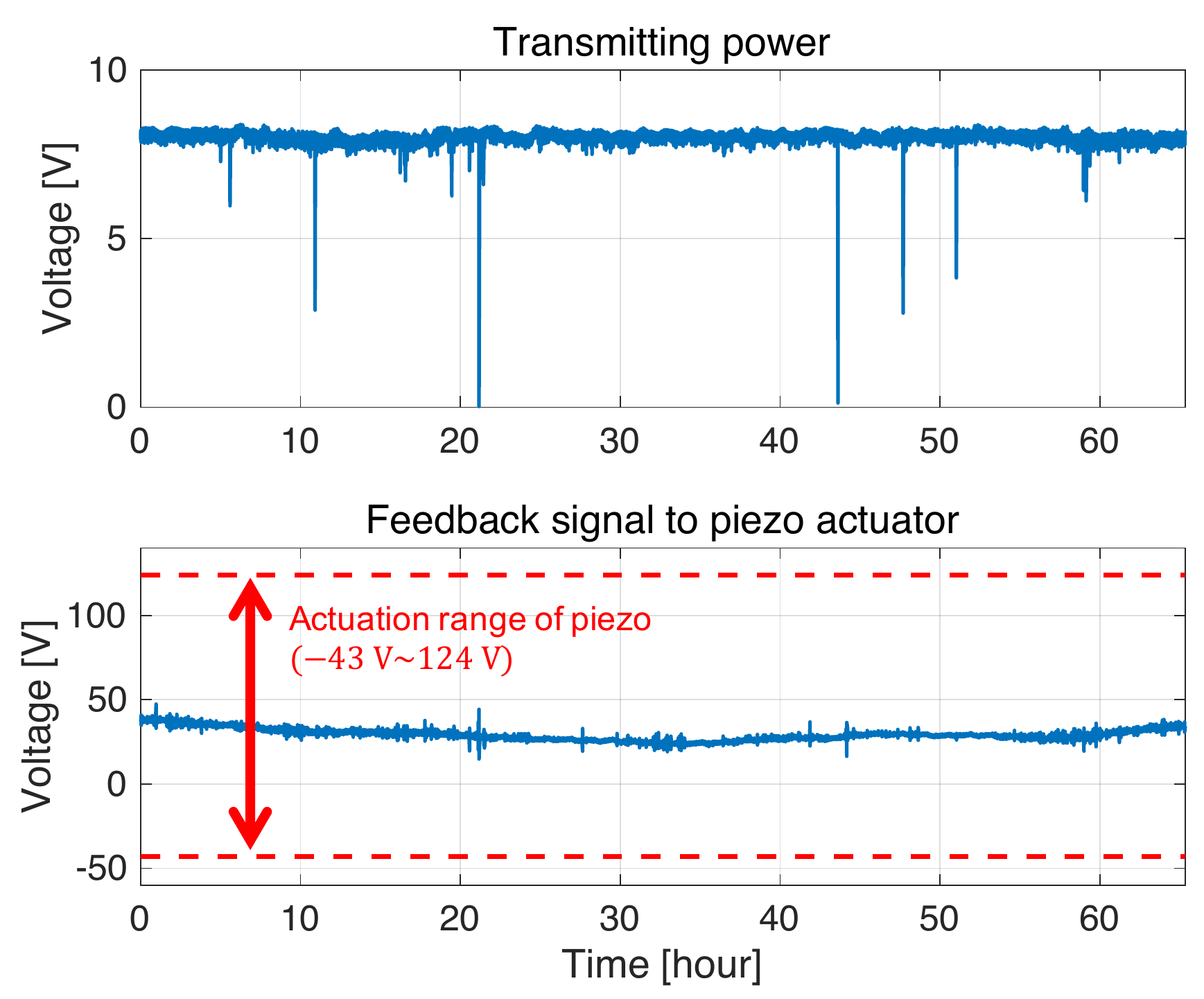}
\end{minipage}
\hspace{30mm}
\begin{minipage}{0.33\linewidth}
\centering
\includegraphics[width=1.32\linewidth]{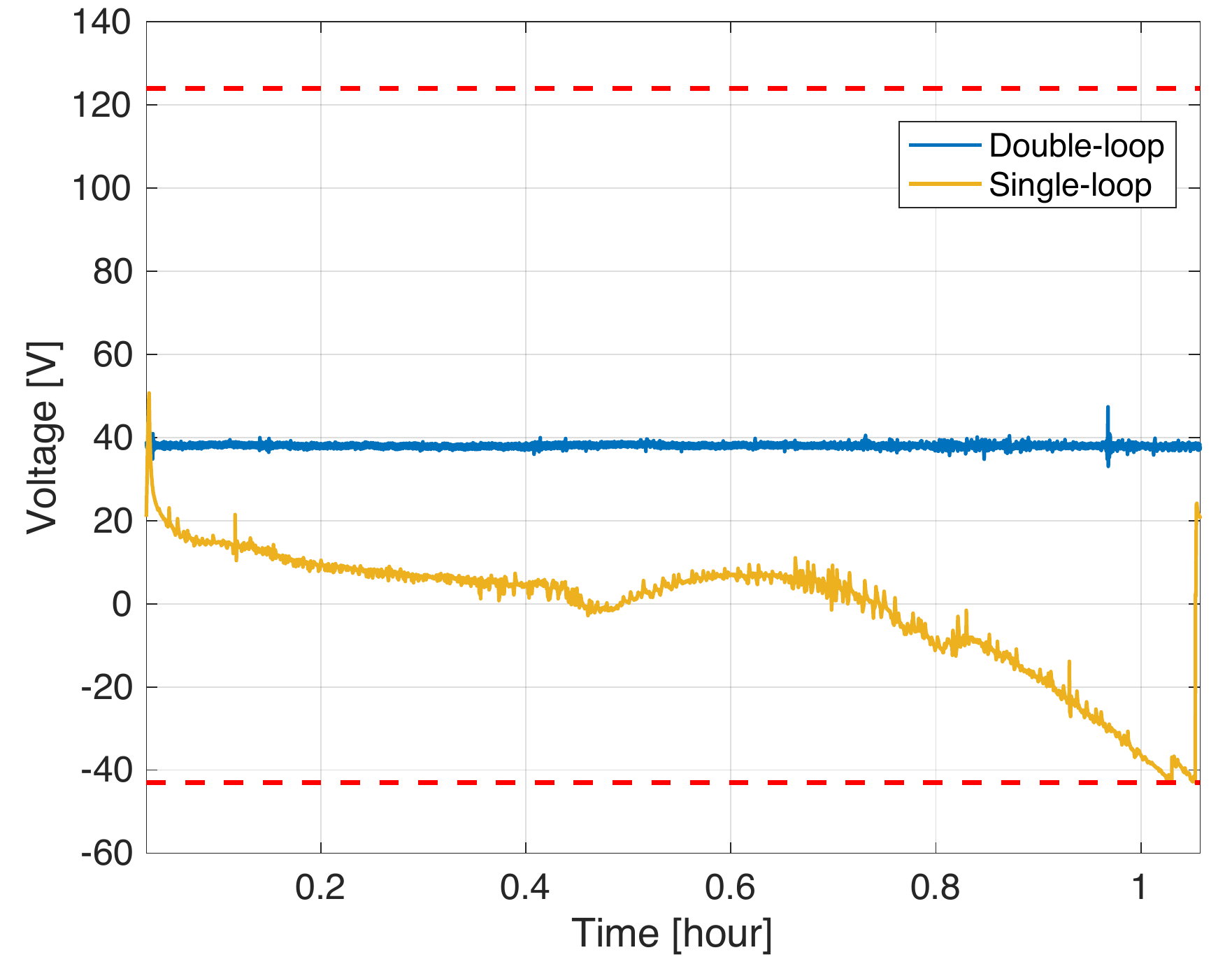}
\end{minipage}
\caption[]{The lock duration time measurement. In the figure of transmitting power, 0.02-second flickers from unknown causes can be seen, although the cavity is not unlocked.
}
\label{fig:double}
\end{figure}

Fig. 2 shows the transmitting light from the cavity and the feedback signal to the piezo actuator when the cavity is locked with the double-loop feedback control system. In this measurement, the lock duration time is improved to more than 60 hours.

The right panel of Fig. 2 shows the feedback signal to the piezo actuator with the single- and double-feedback control for the first one hour of the measurement.
As can be seen from the figure, the large drift in the feedback signal to piezo actuator is decreased greatly due to the double-loop control.

\subsection{Recovery of lock}

To demonstrate the automated cavity locking system, the cavity was intentionally unlocked with vibration, and the transmitting light and the output of the DSP to the temperature of the laser source were measured(Fig. 3).
At the point of 40 seconds in the figure, the cavity is unlocked and the DSP started sweeping the temperature. After around 15 seconds from the unlock, the transmitting light is recovered and the cavity is successfully locked again automatically.

\subsection{Noise of digital signal processor}

To investigate whether the DSP is affecting DANCE Act-1 as a large noise source, the noise of DSP and that of DANCE Act-1 were measured (Fig. 4). 
The displacement noise spectrum of DANCE Act-1 was calculated from the error signal of the ring cavity under feedback control, and the electrical noise of DSP was calibrated to the displacement spectrum noise.
The requirement of DANCE Act-1 was set to $1.4\times 10^{-15}\, \rm{m/\sqrt{Hz}}$ so that the relative intensity noise of the transmitting light can be less than the shot-noise level.
The noise of DSP is less than that of DANCE Act-1 by around two orders of magnitude and also satisfies the requirement of DANCE Act-1.

\begin{figure}
 \begin{minipage}{0.5\hsize}
  \begin{center}
  \hspace*{-1cm}
   \includegraphics[width=0.95\linewidth]{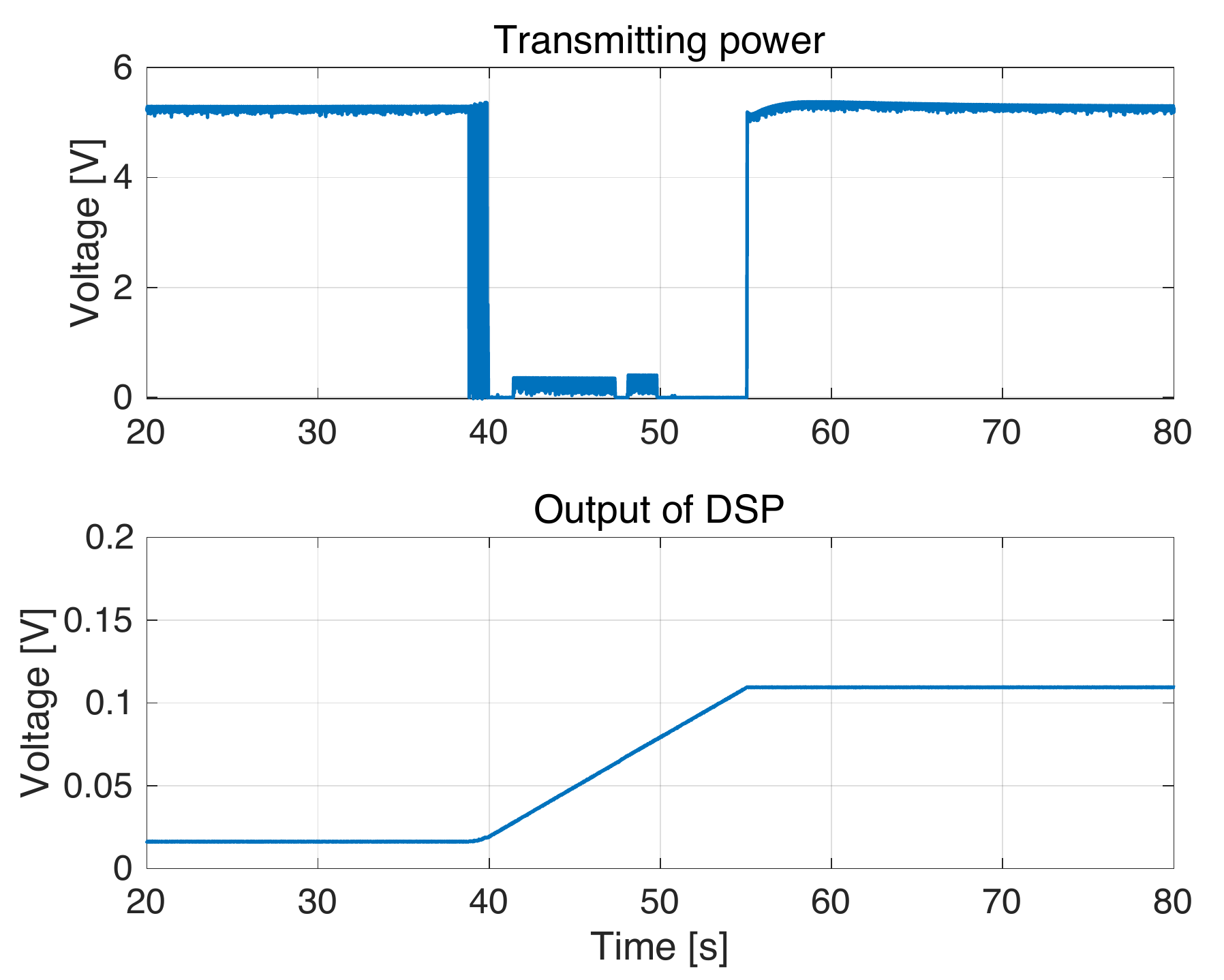}
  \end{center}
\caption[]{The demonstration of the automated cavity locking system.}
  \label{fig:one}
 \end{minipage}
 \hspace{5mm}
 \begin{minipage}{0.5\hsize}
  \begin{center}
   \hspace*{-1.1cm}
   \includegraphics[width=0.95\linewidth]{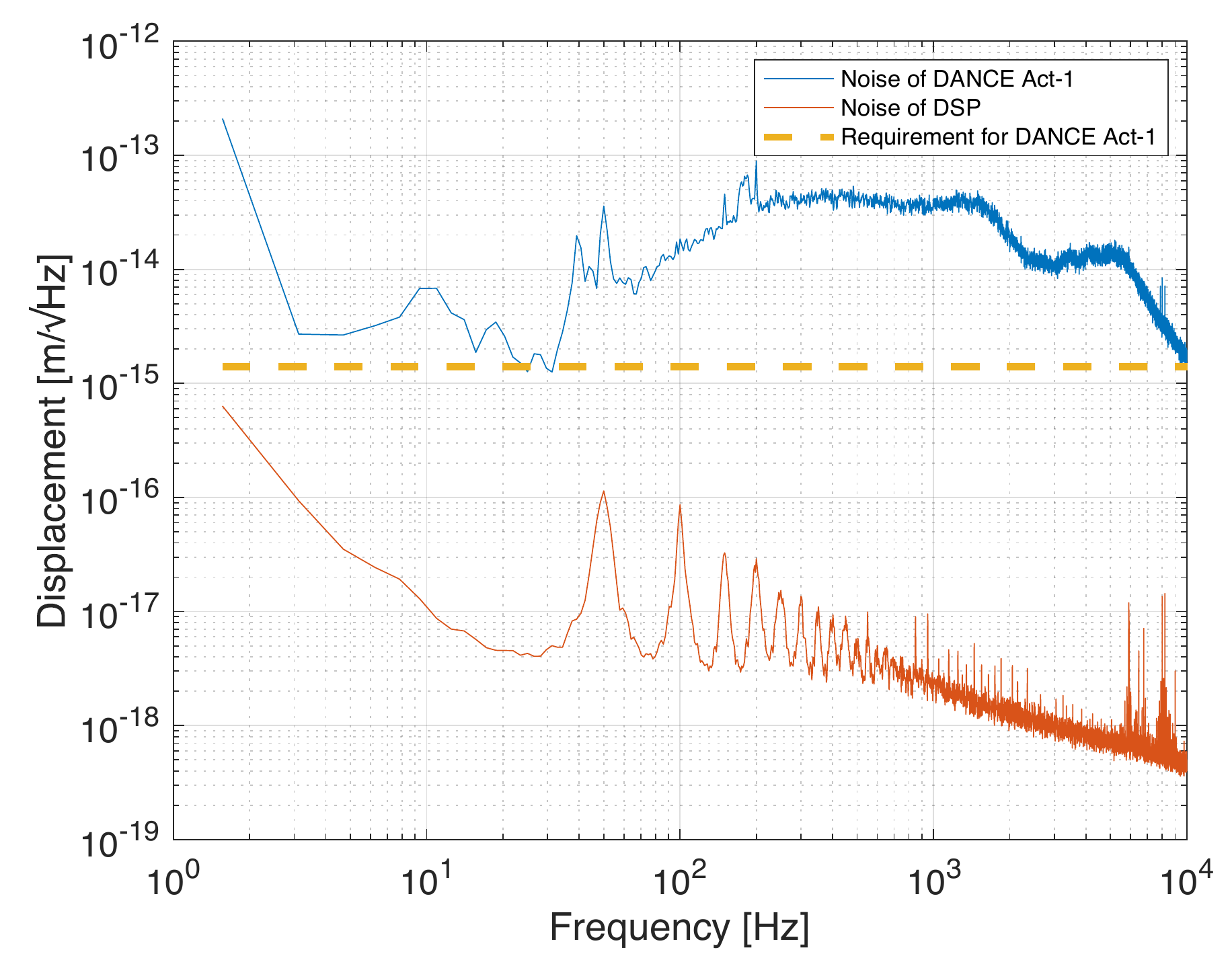}
  \end{center}
\caption[]{The noise spectra of DANCE Act-1 and DSP with feedback control. }
  \label{fig:two}
 \end{minipage}
\end{figure}

\section{Conclusion}

DANCE searches for ALP dark matter with a ring cavity by enhancing the rotation of linear polarization. To realize the long-term observation for better SNR, stable control of the cavity is needed. In this study, the double-loop feedback control system for the longer lock duration time and the automated cavity locking system for dealing with the inevitable unlock were developed. 
As a result, we succeeded in improving the lock duration time to more than 60 hours and in automatic locking of the cavity after around 15 seconds from the unlock.
For future works, we are currently planning to develop an algorithm, which can identify the fastest sweep direction to lock the cavity automatically, and planning to specify the cause of the light flickers observed during long-term lock.

\section*{Acknowledgments}

We would like to thank Shigemi Otsuka and Togo Shimozawa for manufacturing the mechanical parts of the cavity. We would also thank Ooi Ching Pin for correcting English of this document. This work was supported by JSPS KAKENHI Grant Nos. 18H01224, 20H05850, 20H05854 and 20H05859, and JST PRESTO Grant No. JPMJPR200B.

\end{document}